\documentstyle[11pt,aaspp4]{article}
\slugcomment{Submitted to {\it The Astrophysical Journal}}

\begin{document}

\title{The Shapes of Elliptical Galaxies with Central Density Cusps and Massive
Dark Objects}

\author{Barbara S. Ryden,\altaffilmark{1}}
\affil{Department of Astronomy, The Ohio State University, 174 W. 18th
Avenue, Columbus, OH 43210}

\altaffiltext{1}{National Science Foundation Young Investigator;
ryden@astronomy.ohio-state.edu.}

\received{}
\accepted{}

\begin{abstract}
For a sample of 22 core ellipticals, with shallow central cusps,
and 22 power law ellipticals, with steep central cusps, I examine
the apparent axis ratios of isophotes as a function of the
dynamical time at the isophote's mean radius. The two types of
elliptical galaxy have a significantly different distribution of
shapes in the region where the dynamical time is short ($\sim 1 {\,\rm Myr}$);
the core galaxies are rounder in this region. In the outer regions
($t_{\rm dyn} \gtrsim 10 {\,\rm Myr}$),
the two types have indistinguishable shape distributions.
I perform a more detailed analysis of the apparent and intrinsic
shapes at two chosen locations: an inner region, where $t_{\rm dyn} =
0.75 {\,\rm Myr}$, and an outer region, where $t_{\rm dyn} = 50 {\,\rm Myr}$.
The inner isophotes of core galaxies are consistent with their
being randomly oriented oblate objects, with a peak intrinsic
axis ratio of $\gamma = 0.86$. The outer regions of core galaxies
are inconsistent with their being randomly oriented oblate or prolate
objects, at a 99\% confidence level. The oblate hypothesis
cannot be rejected for the inner and outer regions of power law galaxies
at the 90\% confidence level; however, if power law galaxies are
oblate, very few of them can be nearly spherical, with $\gamma > 0.8$.
The observed shapes are consistent with the hypothesis that
the inner regions of core galaxies are nearly spherical and
nearly axisymmetric, perhaps as the result of the physical
process which formed the core.

\end{abstract}


\section{Introduction}

A growing body of evidence (as outlined, for instance, by
\cite{kr95}) indicates that elliptical
galaxies typically harbor a central massive dark object (or MDO).
The masses of MDOs in elliptical galaxies are best determined
for giant ellipticals with nuclear gas disks, such as M87
(\cite{har94}), NGC 4261 (\cite{ffj96}),
and M84 (\cite{bow98}). Stellar dynamics can
also yield estimates for MDO masses, although the estimated
masses are model-dependent. For example,
Magorrian et al.\ (1998) looked at a sample of nearby galaxies
with {\it Hubble Space Telescope (HST)}
photometry and ground-based kinematics.
Of the 32 elliptical galaxies in their sample, they found that
28 were well fitted by an axisymmetric model with a two-integral
distribution function, a constant mass-to-light
ratio, and an optional MDO. Of the galaxies consistent with their model,
26 required the presence of an MDO for the best fit, with a 
typical MDO mass of $M_\bullet \sim 0.005 M_{\rm gal}$, where
$M_{\rm gal}$ is the mass of the stellar component of the galaxy.

In addition to housing central MDOs, most elliptical galaxies
have central cusps in the luminosity density $\nu (r)$.
For a sample of early-type galaxies observed by {\it HST} (\cite{cegIII}),
the logarithmic slope of the luminosity density,
$\eta \equiv - d \log \nu / d \log r$, is significantly different
from zero for most galaxies at $0.1 \arcsec$ from the
galaxy center. Elliptical galaxies fall into two distinct classes,
distinguished by their central density
profile (\cite{jfo94}, \cite{cegI}); ``core'' galaxies have shallow
cusps, with $\eta \sim 0.8$, and ``power law'' galaxies have
steeper cusps, with $\eta \sim 1.9$ (\cite{cegIII}).

The presence of a central MDO or density cusp has a significant
effect on the structure of an elliptical galaxy (\cite{gb85},
\cite{nmv85}). In particular, it determines whether or not
a galaxy can be highly triaxial. (The phrase ``highly triaxial''
can be quantified. Consider an ellipsoid with principal axes
of length $a \geq b \geq c$. The triaxiality is customarily
given by the parameter $T \equiv (1 - b^2 / a^2) / (1 - c^2 / a^2)$.
An ellipsoid with $T \approx 0$ is nearly oblate, an
ellipsoid with $T \approx 1$ is nearly prolate, and an
ellipsoid with $T \approx 1/2$ is highly triaxial.)
In a stellar system with a constant density core,
such as the ``perfect ellipsoid'' (\cite{kuz56}, \cite{dez85}),
stars can follow box orbits, on which stars eventually come
arbitrarily close to the center, or tube orbits, on which
stars circulate about the long or short axis of the system.
Box orbits are generally referred to as the ``backbone'' of
a triaxial galaxy; since tube orbits can only be used to construct
a nearly axisymmetric system, the presence of box orbits is
required for a highly triaxial system. However, in a galaxy
with a central point mass or a steep density cusp, box orbits
are broken up; a star on a box orbit will eventually have a close
encounter with the central mass, subjecting the star to a large-angle
deflection.

Modeling galaxies with density cusps, $\rho \propto r^{-\eta}$,
places restrictions on their permitted equilibrium shapes.
Schwarzschild (1993) examined self-similar triaxial models
with $\eta = 2$; he discovered that self-consistent models
were not possible for extremely flattened galaxies ($c/a = 0.3$).
In his highly flattened triaxial models, Schwarzschild found
that many of the orbits were stochastic, changing shape
significantly over $\sim 160$ orbital times. Merritt \& Fridman (1996)
found that for weak cusps, with $\eta = 1$, as well as
for strong cusps, with $\eta = 2$, a
large fraction of model orbits are stochastic.
The mixing time in phase space for the stochastic
orbits is roughly 100 times the orbital period for a model
with a steep cusp, but may be more than 1000 times
the orbital period for a model with a weak cusp.

Since stochastic orbits in a triaxial potential
are nearly axisymmetric and nearly spherical, it is impossible to
create an equilibrium self-consistent model for a strongly
triaxial galaxy with a central density cusp. For a Jaffe model,
which has $\eta =2$ in the central regions, Merritt (1997)
found self-consistent solutions only for models which were both
nearly axisymmetric ($T \lesssim 0.4$ or $T \gtrsim 0.9$)
and nearly spherical ($c/a \gtrsim 0.8$). Looking at
models with different values of $\eta$, Fridman \& Merritt (1997)
found that cuspy models with $\eta \gtrsim 0.8$ can only have stable
boxlike orbits if they are nearly spherical ($c/a \gtrsim 0.75$).
Studies of scale-free, cuspy disks, with $c/a \to 0$, 
place severe restrictions on the shapes of highly flattened galaxies
(\cite{kui93}; \cite{sz98}); generally speaking, a cuspy
galaxy in equilibrium cannot be both highly flattened and highly
triaxial.

Central point masses are even more effective than central density
cusps at inducing chaotic mixing of orbits. A relatively modest
central black hole, with $M_\bullet / M_{\rm gal} \approx 0.005$, is
as effective as a steep $\eta = 2$ cusp at inducing chaotic
mixing of orbits (\cite{vm98}). Numerical simulations (\cite{mq98})
indicate that a black hole of this mass will change the
galaxy's overall shape on a time scale $\sim 100$ times the orbital
period at the half-mass radius. 

Observed elliptical galaxies have a range of cuspiness, as
expressed by the logarithmic slope $\eta$ of the luminosity
density, and a range of MDO masses, as expressed by the
ratio $x \equiv M_\bullet / M_{\rm gal}$ of the MDO mass
to the total stellar mass of the galaxy. Moreover, cuspiness
and MDO mass appear to be inversely linked. In the sample of early-type
galaxies modeled by Magorrian et al.\ (1998), the 21
core galaxies have a median $x = 0.0072$, and the
11 power law galaxies have a median $x = 0.0033$.
Moreover, three of the power law galaxies, but none of the core
galaxies, are best fit by a model with $x = 0$.
A Kolmogorov-Smirnov (KS) test reveals that the distributions of
$x$ for the two populations differ at the level $P_{\rm KS} = 0.014$.

Core ellipticals and power law ellipticals are known to differ
in many properties other than the slopes of their density cusps.
In addition to having steeper cusps and lower mass MDOs, power
law ellipticals tend to have relatively low luminosity ($M_V > -22$), disky
isophotes, and rapid rotation. Core ellipticals have
relatively high luminosity ($M_V < -20.5$), boxy or elliptical
isophotes, and slow rotation. The two classes also seem to
differ systematically in their apparent axis ratios;
Tremblay \& Merritt (1996) showed that bright
elliptical galaxies (which tend to be core galaxies) are
rounder on average than fainter ellipticals. The shapes
of the fainter ellipticals are consistent with their being
randomly oriented oblate spheroids, but the bright ellipticals
cannot all be oblate.

Tremblay \& Merritt (1996) looked at the average shape of elliptical
galaxies, represented by the luminosity-weighted axis ratio
(\cite{ryd92}). In addition to looking at the average shape,
it is also interesting to examine how the apparent shape of
a galaxy's isophotes changes with radius. For instance, consider
an elliptical galaxy which is initially highly triaxial. (This
is not an absurd assumption; the merger of two disk galaxies,
for example, is capable of producing a highly triaxial remnant
[\cite{bar92}].) The central cusp and MDO of the elliptical galaxy
will make it more nearly axisymmetric and more nearly spherical,
starting from the inner regions, where the dynamical time is shortest,
and working steadily outward. If the characteristic time for
changing the shape is $\sim 100 t_{\rm dyn}$, then
at a given galaxy age $t_{\rm gal}$, the inner region where
$t_{\rm dyn} < t_{\rm gal}/100$ will be nearly axisymmetric
and nearly spherical, while the outer regions will still
retain their initial shape. 

Is this naive picture true? Are real elliptical galaxies
more nearly axisymmetric and spherical in their inner regions
than in their outer regions? Does the variation of shape with
radius differ for core galaxies and for power law galaxies?
To investigate these questions, I examine the apparent shapes,
as a function of isophotal radius, for a sample of nearby elliptical galaxies.
The details of the galaxy sample used are given in Section 2 below.
In section 3, I compare the shapes of the core galaxies
and of the power law galaxies in the sample. For each galaxy,
I compute the axis ratio of the isophote where the dynamical
time is $t_{\rm dyn} \sim $ 0.75 Myr, and that of the
isophote is $t_{\rm dyn} \sim $ 50 Myr. Comparing the
inner isophotes, where the dynamical time is short, I find
that the core galaxies are significantly rounder than
the power law galaxies. At the outer isophotes, the core
galaxies and power law galaxies have the same distribution of
apparent shapes. In section 4, I summarize the results and
their implications.

\section{Galaxy Sample}

The sample consists of 44 elliptical galaxies drawn from the
study of Faber et al.\ (1997; hereafter F97), who compiled early-type galaxies
that had been imaged by the {\it HST} Planetary Camera through
filter F555W (approximately the $V$ band).
My sample of 44 ellipticals excludes NGC 7768,
but includes every other galaxy in the F97 compilation
which is classified as an E galaxy in the
{\it Second Reference Catalog of Bright
Galaxies} (de Vaucouleurs et al. 1976).
NGC 7768 was excluded from the sample because its central
dust disk strongly affects the surface brightness at $V$
within about $0.4\arcsec$ of the galaxy's center (Grillmair et al. 1994).

Although NGC 7768 was the only galaxy discarded from
the sample due to the presence of dust, many of the remaining
ellipticals contain detectable amounts of dust in their
central regions (\cite{vdf95}), usually within a kiloparsec
or so of the galaxy's center.
Thus, the effect of dust on isophote shapes is a potential
concern when determining the shapes of the central regions
of ellipticals. Some information on the effect of dust on
the central isophote shapes comes from the study of Carollo
et al. (1997); during their study of elliptical galaxies with
kinematically distinct cores, they obtained {\it HST} WFPC2 images
at F555W and F814W (roughly $V$ and $I$) for 5 of the elliptical
galaxies in the F97 sample: NGC 1700, NGC 3608, NGC 4365,
NGC 4552, and NGC 5813. From their $V$ and $I$ images, Carollo et al.
created ``dust-improved'' images that minimized, but did not
eliminate, the effects of dust absorption. A comparison of
the isophote parameters of the dust-improved images of Carollo
et al. (1997) with the ``unimproved'' $V$ images used by Faber et al.\ (1997)
indicates no systematic difference between the isophote axis ratios
in the central $2.5\arcsec$ of the images (\cite{cfi97}).
Speaking more generally, observed circumnuclear dust features
in early-type galaxies are usually patchy, asymmetric, and/or misaligned
with the galaxy's major axis (\cite{vdf95}). Features of this sort
cause significant deviations from pure elliptical isophotes, as
measured by the $3\theta$ and $4\theta$ terms in the Fourier
expansion of the isophotes (\cite{pdi90}), but do not strongly affect the
measured axis ratio of isophotes. Because of the relative insensitivity
of axis ratios to the presence of dust, I will use the published
isophotal data for the F97 galaxies at face value, making
no corrections for dust.

Over two-thirds of the galaxies in the F97 compilation
are taken from the study of Lauer et al. (1995; hereafter L95).
Lauer et al. were concerned with the central structure of nearby
early-type galaxies. The elliptical galaxies in their sample
span a broad range of luminosity, from the brightest cluster galaxy
NGC 6166 ($M_V = -23.5$) to the M32 analog VCC 1199 ($M_V = -15.2$).
The majority of the galaxies in the L95 sample were selected because
ground-based photometry failed to resolve a central core. The L95
sample, and hence the F97 compilation, is weighted toward
galaxies with high central surface brightness. The results which
I derive in the present paper, therefore, are inapplicable to
low surface brightness galaxies.

Of the 44 elliptical galaxies in the F97 sample, 22 are classified as core
galaxies and 22 are classified as power law galaxies.
The core subsample is listed in Table 1, and the power law
subsample is listed in Table 2. For 35 of the galaxies, the isophotal
parameters derived from the {\it HST} data
are given by Lauer et al.\ (1995); these galaxies are indicated by
the label `L95' in the `inner isophote' column of Tables 1 and 2.
For 5 of the galaxies, the isophotal parameters are given by van den
Bosch (1994); these are indicated by the label `V94' in Tables
1 and 2. The isophotal parameters for NGC 4486 [M87] are given
in Lauer et al.\ (1992; hereafter L92) and the parameters for NGC 221 [M32]
are given by Lauer et al. (1998; hereafter L98).
For NGC 1600, NGC 3379, NGC 4552,
NGC 4621, and NGC 4649, isophotal parameters were 
provided in advance of publication (\cite{lau99}; hereafter L99). 

The {\it HST} photometry for the galaxies in the sample includes roughly the
central 10 arcsec of each galaxy, equivalent, at the distance of
the Virgo cluster, to a linear scale of about 800 parsecs and a dynamical
time, for a bright elliptical galaxy, of a few million years.
To find the shapes of the outer
regions of the galaxies, where the dynamical time is longer, I used
isophotal parameters drawn from the published literature.
For 17 galaxies, the outer isophotal parameters were taken from
the work of Capaccioli and collaborators (\cite{cpr88}, hereafter C88;
\cite{ccr90}, hereafter C90; \cite{ccd94}, hereafter C94). These
studies (C88, C90, \& C94) combine CCD and photographic photometry
in the $B$ band to derive ellipticity and luminosity profiles to a
limiting surface brightness of $\mu_B \sim 28 {\,\rm mag} {\,\rm arcsec}^{-2}$;
the ground-based $B$ profiles were matched to the {\it HST} $V$
profiles assuming a constant $B-V$ color outside the region imaged
by {\it HST}. For an additional 12 galaxies, the outer isophotal
parameters were taken from Peletier et al. (1990), who obtained
CCD surface photometry for a selection of nearby bright ellipticals;
they derived ellipticity and luminosity profiles in the $R$ band
and luminosity profiles in the $B$ band to a limiting surface
brightness of $\mu_B \sim 26 {\,\rm mag} {\,\rm arcsec}^{-2}$.
I computed an approximate $V$ band profile by setting $V = (B+R)/2$,
then adjusted the zero point to match the outermost isophote of
the published {\it HST} profile. Since the observed axis ratios of
elliptical galaxies do not vary significantly with color (\cite{pdi90}),
I used the $R$ band ellipticities as my measure of outer isophote shape.
I drew additional ground-based data from other published sources,
as listed in the ``Outer isophotes'' column of Tables 1 and 2.
When the ground-based photometry was not in the $V$ band, I assumed
that the galaxy's color in the outer regions was constant, with a
value equal to the color at the outermost published {\it HST} isophote.
I also assumed that the isophote axis ratio was not a function of
color.

Given the somewhat eclectic nature of the F97 sample, it is prudent
to check that the shapes of the F97 ellipticals are representative
of elliptical galaxies as a whole, and are not biased toward nearly
round or highly flattened shapes. To find an overall average shape
for each elliptical, I compute the luminosity-weighted mean axis
ratio $\overline{q}$, as detailed in Ryden (1992). The value of
$\overline{q}$ was computed within the isophote at which $\mu_V =
19.80 {\rm\,mag}{\rm\,arcsec}^{-2}$, corresponding, for a galaxy
with $B - V = 0.95$, to the surface brightness $\mu_B = 20.75
{\rm\,mag}{\rm\,arcsec}^{-2}$ used to define the fiducial diameter
$D_n$ (Dressler et al. 1987). Defined in this manner, the mean and
standard deviation of $\overline{q}$ for the 44 ellipticals in the
F97 sample are $0.796 \pm 0.114$. I compute $\overline{q}$ in the
identical manner for a sample of 165 bright nearby ellipticals observed
by Djorgovski (1985), 26 of which are also in the F97 compilation.
The mean and standard deviation of $\overline{q}$ for the galaxies
in the Djorgovski (1985) sample are $0.803 \pm 0.103$. A KS test
comparing the distribution of $\overline{q}$ for the two samples indicates
that they are statistically indistinguishable, with $P_{\rm KS} = 0.67$.
Other studies in which the mean axis ratio $\overline{q}$ is defined in
different ways, yield distributions of $\overline{q}$ which are similar
to those of the F97 and Djorgovski (1985) sample: a distribution peaking
at $\overline{q} \sim 0.8$, with relatively few galaxies at $\overline{q}
\sim 1$ (\cite{bg80}; \cite{fv91}; \cite{fid91}; \cite{lml92}). Thus,
the F97 is not obviously biased toward round or flattened ellipticals.

By combining the {\it HST} isophotes with the ground-based isophotes, where
available, I had, for each galaxy, a knowledge of the isophotal axis
ratio $q \equiv b/a$ and surface brightness $\mu$ as a function of
the semimajor axis $a$. However, from a physical point of
view, it is more useful to label each isophote with its dynamical
time $t_{\rm dyn}$ than with its
value of $a$ in arcseconds or in parsecs. The definition I use for
dynamical time is (Binney \& Tremaine 1987)
\begin{equation}
t_{\rm dyn} = {\pi \over 2} \left( {G [ M (r) + M_\bullet ] \over r^3 } \right)^{-1/2} \ ,
\end{equation}
where $M(r)$ is the stellar mass within a sphere of radius $r$ and
$M_\bullet$ is the mass of the central MDO (if any). The dynamical
time defined in this way is 1/4 the orbital time of a star on a circular
orbit with radius r.

To find the spherically averaged luminosity distribution of each galaxy,
I start with the surface brightness
profile $\mu (R)$, where $R \equiv q^{1/2} a$, deconvolve the
surface brightness to find the luminosity density $\nu (r)$, then
integrate to find the luminosity $L (r)$.
The transition from the luminosity profile $L(r)$ to the mass
profile $M(r)$ was made using the assumption of a constant
mass-to-light ratio $\Upsilon$. For 15 of the 22 core galaxies
and 5 of the 22 power law galaxies, a best-fitting value of
$\Upsilon$ and a best-fitting value of $M_\bullet$
was calculated by Magorrian et al.\ (1998); for these
galaxies, I use the best-fitting Magorrian et al. values for $\Upsilon$
and $M_\bullet$.
To the remaining galaxies, unfitted by Magorrian et al.,
I assigned a mass-to-light ratio $\Upsilon$ according
to the relation
\begin{equation}
\log ( \Upsilon / \Upsilon_\odot ) = -1.11 + 0.18 \log ( L_{\rm gal} / L_\odot ) \ ,
\end{equation}
which Magorrian et al.\ (1998) found to be the formal best-fitting
straight line to $( \log L_{\rm gal} , \log \Upsilon )$ for the early-type
galaxies in their sample.
There remains the problem of assigning a value of $M_\bullet$ to
those galaxies in my sample that were not kinematically modeled
by Magorrian et al.\ (1998). For both power law galaxies and core
galaxies Magorrian (1998) found that the parameter $x \equiv
M_\bullet / M_{\rm gal}$ was well fit by a Gaussian in $\log x$.
For core galaxies, the mean and standard deviation in $\log x$ were
$(-2.13, 0.31)$; for power law galaxies, the mean and standard
deviation were $(-2.57, 0.55)$. For the purposes of computing
approximate dynamical times, I assigned an MDO mass of $x =
10^{-2.13} = 0.0074$ to core galaxies not modeled by Magorrian et al.
and a mass of $x = 10^{-2.57} = 0.0027$ to power law galaxies.

The approximate value of $t_{\rm dyn} (r)$ found by the method outlined
above is, of course, laden with errors. I assume spherical symmetry;
this is a minor source of error. I also assume that the value of $\Upsilon$
found by Magorrian et al. (1998) for the central regions of galaxies
can be safely used in the outer regions as well. If the mass-to-light
ratio actually increases with radius, then my estimates of the
dynamical time will actually be overestimates. I also ignore the
observed spread in $\log x$ for each class of galaxy when
assigning values of $x$ to the ellipticals without kinematic
information. The uncertainty in $x$ leads to significant uncertainty in
the dynamical time only within the radius $r_{\rm eq}$ at which
$M(r) = M_\bullet$. For the 5 power law ellipticals fitted by Magorrian
et al. (1998) the median value of $r_{\rm eq}$ was only 8 pc, with
a corresponding dynamical time of $t_{\rm eq} \approx 0.04 {\rm\,Mpc}$.
For the 15 core galaxies fitted, the median value of $r_{\rm eq}$
was 230 pc, with a corresponding dynamical time of $t_{\rm eq} \approx
0.9 {\rm\,Mpc}$.
The standard deviation in $\log x$ of 0.31 for core galaxies corresponds
to an uncertainty of a factor 2 in $M_\bullet$. In the region where
$r \ll r_{\rm eq} \approx 230 {\rm\,pc}$, this leads
to a 43\% uncertainty in $t_{\rm dyn}$ for
a given value of $r$, or equivalently, a 27\% uncertainty in $r$ for
a fixed value of $t_{\rm dyn}$.

One last cautionary note about MDO masses: the values of $M_\bullet$
deduced by Magorrian et al. (1998) are dependent on the two-integral
models which they assume for the kinematics. If the velocity dispersion
is allowed to be radially anisotropic, then the central MDO mass required
will be reduced. The limited information available about velocity
anisotropy in the centers of ellipticals, as reviewed by van der Marel
(1998), is consistent with core galaxies being radially anisotropic
and power law galaxies being roughly isotropic, but the uncertainties
are still great. If, in fact, the Magorrian et al. (1998) masses
are overestimates, then the values of $t_{\rm dyn}$ computed
in this paper will be too short within the region inside $r_{\rm eq}$.
Outside $r_{\rm eq}$, the dynamical times are not greatly affected
by the MDO mass assumed.

For each galaxy in my sample, I compute $t_{\rm dyn}$ as a function
of semimajor axis, in the range $0.1\arcsec < a < a_{\rm max}$,
where $a_{\rm max}$ is the semimajor axis of the outermost
isophote in the ground-based photometry. The minimum value
of $a = 0.1\arcsec$ corresponds to the resolution limit
of the {\it HST} Planetary Camera images. For the 22 core galaxies
in the sample, the median dynamical time at $a = 0.1\arcsec$ is
0.03 Myr; for the 22 power law galaxies, the median dynamical time
at $a = 0.1\arcsec$ is 0.08 Myr. For the 22 core galaxies, the
median dynamical time at $a = a_{\rm max}$ is 150 Myr; for the
22 power law galaxies, the median dynamical time at $a = a_{\rm max}$
is 60 Myr. Thus, when making statistical comparisons between
the core galaxies and power law galaxies, the statistics are
best in the range $0.08 {\,\rm Myr} \lesssim t_{\rm dyn} \lesssim
60 {\,\rm Myr}$. (The more restricted range of dynamical times for which
{\it all} galaxies in the sample have isophotal data is $0.3 {\,\rm Myr}
< t_{\rm dyn} < 8 {\,\rm Myr}$.)

\section{Analysis}

Once each isophote has a dynamical time $t_{\rm dyn}$ and apparent
axis ratio $q$ associated with it, I can examine the 
distribution of $q$ at fixed values of $t_{\rm dyn}$ for
selected subsamples of ellipticals.
As a first step in the analysis,
Figure 1 shows the mean value of $q$ (upper lines) and the standard
deviation of $q$ (lower lines) as a function of dynamical time.
The solid lines indicate the mean and standard deviation for
the core galaxies; the dashed lines are the values for the power law galaxies.
For a wide range of dynamical times, $0.1 {\,\rm Myr} \lesssim t_{\rm dyn}
\lesssim 50 {\,\rm Myr}$, the core galaxies are rounder on average
than the power law galaxies, and have a narrower distribution of $q$.
Moreover, though there is no systematic trend in the average value
of $q$ with $t_{\rm dyn}$ for the power law galaxies, the core
galaxies become steadily flatter on average for $t_{\rm dyn} \gtrsim
2 {\,\rm Myr}$ and for $t_{\rm dyn} \lesssim 0.2 {\,\rm Myr}$.

Are the differences in projected shape between core galaxies
and power law galaxies statistically significant? After all, there
are at most 22 galaxies of each type at a given $t_{\rm dyn}$,
and 22 is not a huge number. To test whether the shapes of
core galaxies and power law galaxies are drawn from the same
parent distribution, I performed KS tests. The results of the
comparison between the distribution of $q$ for core galaxies
and for power law galaxies is shown in Figure 2. The KS probability
is $P_{\rm KS} < 0.01$ only in the limited range
$0.48 {\,\rm Myr} < t_{\rm dyn} < 3.3 {\,\rm Myr}$;
within this range, the shapes of the core galaxies in my sample are
different, at the 99\% confidence level, from those
of the power law galaxies. Both in the central regions of
galaxies, where $t_{\rm dyn} \lesssim 0.2 {\,\rm Myr}$, and
in the outer regions, where $t_{\rm dyn} \gtrsim 20 {\,\rm Myr}$,
the shape distribution for core galaxies is statistically
indistinguishable, at the 90\% confidence level, from that
of power law galaxies.

The puzzling question is why the intermediate region of
core ellipticals, where $t_{\rm dyn} \sim 1 {\,\rm Myr}$,
should be rounder on average than both the central region, where $t_{\rm dyn}
\lesssim 0.2 {\,\rm Myr}$, and the outer region, where
$t_{\rm dyn} \gtrsim 20 {\,\rm Myr}$. The shape of the
central regions is unfortunately complicated, in many
cases, by the presence of substructure. Among the 44 ellipticals
I am examining, the radius at which
$t_{\rm dyn} = 0.2{\,\rm Myr}$ ranges from about 8 parsecs
for the low luminosity elliptical VCC 1199 to 150 parsecs for
the brightest cluster galaxy NGC 6166, with a typical value of
a few tens of parsecs.
This is the length scale on which
elliptical galaxies frequently display signs of having
stellar disks (\cite{vfj94}, \cite{cegIV}), dust disks
and rings (\cite{vdf95}), stellar nuclei (\cite{cegI}),
and kinematically distinct cores. Due to the relative complexity
of the central regions of ellipticals, I will not
analyze in detail their shapes in this paper, concentrating
on the intermediate ($t_{\rm dyn} \sim 1 {\,\rm Myr}$) and
outer ($t_{\rm dyn} \gtrsim 20 {\,\rm Myr}$) regions.

Although the mean and standard deviation of $q$, as plotted
in Figure 1, are valuable bits of information, they don't
provide all the information contained in $f(q)$, the full
distribution of apparent shapes at a given $t_{\rm dyn}$.
Since a full knowledge of $f(q)$ provides important constraints
on the permissible range of intrinsic shapes, I will examine
in detail the distribution of $q$ for isophotes at two
selected dynamical times. The inner isophote will lie in
the regime $0.48 {\,\rm Myr} < t_{\rm dyn} < 3.3 {\,\rm Myr}$,
where the shapes of the core and power law ellipticals are
significantly different, and the outer isophote in the regime $t_{\rm dyn}
> 20 {\,\rm Myr}$, where their shapes are statistically
indistinguishable.

For the shorter dynamical time, and inner isophote, I choose
$t_{\rm dyn} = 0.75 {\,\rm Myr}$.
The axis ratio $q_{\rm in}$ is then
defined as the observed axis ratio of the isophote with
semimajor axis
$a = r_{\rm in}$, where $t_{\rm dyn} ( r_{\rm in} ) = 0.75 \,{\rm Myr}$.
In angular units, the values of $r_{\rm in}$ for the sample
range from 0.16 arcsec for Abell 1020 to 14 arcsec for the
nearby compact elliptical NGC 221.
In physical units, the values of $r_{\rm in}$ range from
32 pc for VCC 1440 to 330 pc for NGC 6166.

For core ellipticals, a uniquely interesting length scale is the
break radius at which the slope of the surface brightness profile
changes. Faber et al. (1997) computed a break radius $r_b$ by
fitting a ``nuker'' law (\cite{cegI}),
\begin{equation}
\mu(r) = 2^{(\beta-\gamma)/\alpha} \mu_b \left( {r \over r_b} \right)^\gamma
\left[ 1 + \left( {r \over r_b} \right)^\alpha \right]^{(\gamma-\beta)/\alpha}
\ .
\end{equation}
The dynamical time at the break radius varies from one core galaxy
to another; for the core ellipticals in the F97 sample, the value of
$t_{\rm dyn} (r_b)$ ranges from $0.03 {\rm\,Myr}$ for NGC 4486B
to $4.6 {\rm\,Myr}$ for NGC 6166, with a median value of $0.9 {\rm\,Myr}$.
For most of the core ellipticals in the sample, $r_{\rm in}$, as
defined above, is comparable to $r_b$; the values of $r_{\rm in}$
range from $0.24 r_b$ for NGC 1600 to $9.6 r_b$ for NGC 4486B, with a
median value of $r_{\rm in} = 0.7 r_b$.

For the longer dynamical time, and outer isophote, I choose
$t_{\rm dyn} = 50 {\,\rm Myr}$.
The axis ratio $q_{\rm out}$
is then defined as the observed axis ratio of the isophote
with $a = r_{\rm out}$, where $t_{\rm dyn} ( r_{\rm out} ) = 50 \,{\rm Myr}$.
For 2 of the core galaxies and 7 of the power law galaxies
in the sample, $r_{\rm out}$ lies outside the region where published
surface photometry is available.
For the galaxies in the sample
with photometry available out to $r_{\rm out}$, the angular size
of $r_{\rm out}$ ranges from 8.7 arcsec for Abell 1831 to 210 arcsec
for NGC 4472. In physical units, the values of $r_{\rm out}$ range
from 3.2 kpc for NGC 4458 to 16 kpc for NGC 1600. For the core
galaxies, $r_{\rm out}$ always lies well outside the break radius;
the values of $r_{\rm out}$ range from $10 r_b$ for NGC 4874 to $230 r_b$
for NGC 3608.

Values of $q_{\rm in}$ and $q_{\rm out}$ for the core galaxies
in the sample are given in Table 1; values of $q_{\rm in}$ and
$q_{\rm out}$ for the power law galaxies are given in Table 2.
The cumulative distribution functions are displayed graphically
in Figure 3. The upper panel shows the distribution of $q_{\rm in}$
and $q_{\rm out}$ for the core galaxies; the lower panel
shows the equivalent distributions for the power law galaxies.
In each panel, the solid line gives the distribution of $q_{\rm in}$,
and the dotted line gives the distribution
of $q_{\rm out}$. A glance
at Figure 3 shows that $q_{\rm in}$ for the core galaxies
has a significantly different distribution from the other three
populations. This visual impression is confirmed by a Kolmogorov-Smirnov
test applied to the different distributions. The results of the
KS test are displayed in Table 3.
The only pairs of distributions which
differ at the 99\% confidence level, as measured by the KS test,
are $q_{\rm in}$ for core galaxies and $q_{\rm out}$
for core galaxies, with $P_{\rm KS} = 0.0059$, and $q_{\rm in}$
for core galaxies and $q_{\rm in}$ for power law galaxies,
with $P_{\rm KS} = 0.0029$.

Within the subsample of core ellipticals, $q_{\rm in}$ is uncorrelated
with the absolute magnitude $M_V$ of the galaxy; a similar
lack of correlation between $q_{\rm in}$ and $M_V$ is seen within
the subsample of power law ellipticals. This indicates that the difference
in $q_{\rm in}$ between core and power law ellipticals is not simply
a luminosity-dependent effect, reflecting the greater average
luminosity of core galaxies, but rather is more fundamentally
linked to the physical difference between core and power law galaxies.
(This result also suggests that the finding of Tremblay \& Merritt
[1996] -- that bright ellipticals ($M_B < -20$) are rounder than
faint ellipticals -- is based on the fact that core ellipticals are rounder
than power law ellipticals in their inner regions. The luminosity-weighted
mean axis ratio used by Tremblay \& Merritt is strongly weighted
to the bright inner regions of the galaxies, where the core ellipticals,
which dominate the $M_B < -20$ sample, are rounder than power law ellipticals.)

Instead of looking at the cumulative distribution function,
$F(q)$, for a sample of $N$ axis ratios, I can compute a
nonparametric kernel estimate of the distribution function
$f(q)$, defined so that $f(q)dq$ is the fraction of all galaxies
with axis ratios in the range $q \to q+dq$. A full discussion
of using nonparametric estimates for $f(q)$ is given by
Tremblay \& Merritt (1995) and Ryden (1996); I will give
only a brief outline here.
Given a sample $q_1$, $q_2$, $\dots$ , $q_N$ of apparent axis
ratios, the kernel estimate of $f(q)$ is
\begin{equation}
{\hat f}(q) = {1 \over N h} \sum_{i=1}^N K \left( {q - q_i \over h}
\right) \ ,
\end{equation}
where $K$ is the kernel function. To produce a smooth differentiable
estimate, I used a Gaussian function,
\begin{equation}
K (x) = {1 \over \sqrt{2 \pi}} e^{-x^2/2} \ ,
\end{equation}
of width $h = 0.9 \sigma N^{-0.2}$, where $\sigma$ is the
standard deviation of the sample. This kernel width, for distributions
that are not strongly skewed, minimizes the integrated mean square
error (\cite{vfl94}). Because $q$ is limited to the range $0
\leq q \leq 1$, I use reflective boundary conditions at $q = 0$
and $q = 1$ to ensure the proper normalization for ${\hat f}(q)$. Because
these boundary conditions are artificially imposed, the prudent
reader will not place much confidence in the shape of ${\hat f}$ within
a distance $\sim h$ of the boundaries at $q=0$ and $q=1$.

Error intervals can be placed on the estimated $\hat f$ by
a process of bootstrap resampling (\cite{mt94}).
From the original data set $q_1$, $q_2$, $\dots$ , $q_N$, I draw,
with replacement, a new set of $N$ data points and compute a
new estimate of $\hat f$. After making 1000 bootstrap estimates,
I use them to place error intervals on the original estimate $\hat f$.
For instance, the 80\% confidence interval at a given value of
$q$ is the range in $\hat f$ such that 10\% of the bootstrap
estimates lie above the interval and 10\% of the bootstrap
estimates lie below the interval. 

The upper panel of Figure 4 shows the distribution of apparent
shapes for the inner isophotes of core galaxies,
where $t_{\rm dyn} = 0.75 {\,\rm Myr}$. The
solid line is the best estimate for ${\hat f} ( q_{\rm in} )$;
the dashed lines give the 80\% error interval and the dotted
lines give the 98\% error interval. The peak in $\hat f$ is
at $q_{\rm in} = 0.91$ (a nearly round shape, equivalent
to an E1 galaxy). The corresponding
distribution for the shapes of the outer isophotes of
core galaxies is given in the upper panel of Figure 5; note
that the peak in $\hat f$ for the outer regions is at
$q_{\rm out} = 0.76$.
The distribution of apparent shapes for the inner regions of power law
galaxies is shown in the upper panel of Figure 6. The corresponding
distribution of shapes for the outer regions of power law
galaxies is shown in the upper panel of Figure 7. Note that
the distribution $\hat f$ peaks at $q_{\rm in} \approx 0.7$ for
power law galaxies, in both their inner and outer regions.

So far, I have dealt only with the apparent projected shapes
of galaxies, and not with their intrinsic three-dimensional shapes.
For an individual galaxy, it is impossible to determine the
intrinsic shape merely from a knowledge of the projected
photometry. However,
if I assume that the galaxies in a sample are all randomly
oriented oblate spheroids, I can perform a mathematical
inversion of $\hat f (q)$ to find the underlying distribution
${\hat f}_o (\gamma)$ of intrinsic shapes, where $\gamma = c/a$
is the intrinsic axis ratio of the oblate spheroid, with
$0 \leq \gamma \leq 1$. Similarly, if I assume that the
galaxies are randomly oriented {\it prolate} spheroids, I
can perform an inversion to find ${\hat f}_p (\gamma)$, the
underlying distribution of axis ratios for the prolate galaxies.

The game of finding the distribution of intrinsic shapes of elliptical
galaxies, using the oblate and prolate hypotheses, has been played
for several decades (\cite{hub26}; \cite{sfs70}; \cite{bdv81}).
The use of kernel estimates and bootstrap resampling permits
us to reject or accept the oblate hypothesis (or the prolate
hypothesis) at a given level of statistical confidence. For
instance, I can take a given sample of axis ratios, and invert
the best estimate $\hat f$ and the estimate for each of the 1000
bootstrap resamplings, using the hypothesis that the galaxies
are randomly oriented oblate spheroids. This gives me a
best estimate for ${\hat f}_o (\gamma)$ as well as 1000 bootstrap estimates.
At each value of $\gamma$, an upper limit can be placed on ${\hat f}_o$
at the 90\% confidence level, for example, by finding the value of
${\hat f}_o$ such that 10\% of the bootstrap estimates fall above
this value. If this upper confidence level falls below zero, then
the oblate hypothesis can be rejected at the 90\% (one-sided) confidence
level, since negative values of ${\hat f}_o$ are unphysical. Similar
analysis of ${\hat f}_p$ can be used to reject or accept the
prolate hypothesis.

For the inner regions of core galaxies, where $t_{\rm dyn} = 0.75
{\,\rm Myr}$, the nonparametric estimate
${\hat f}_o$, given the oblate hypothesis, is shown in the middle panel of
Figure 4. The best fit (shown as the solid line) dips below zero for
$\gamma$ close to one, but given the width of the error intervals,
the oblate hypothesis cannot be rejected at the 90\% confidence level.
The best fitting ${\hat f}_o$ peaks at $\gamma = 0.86$, with a hint of
a secondary peak at $\gamma = 0.68$. The prolate hypothesis, illustrated
in the bottom panel of Figure 4, is statistically acceptable for
the inner region of core galaxies; ${\hat f}_p$ peaks
at $\gamma = 0.87$.

For the outer regions of core galaxies, where $t_{\rm dyn} = 50
{\,\rm Myr}$, the best estimate for
${\hat f}_o$ is shown in the middle panel of Figure 5.
The best fit, shown as the solid line, dips below zero
for $\gamma > 0.83$, and the upper bound of the 98\% confidence
interval is below zero for $\gamma > 0.932$. Thus, the oblate
hypothesis can be rejected at the 99\% one-sided confidence level.
The prolate hypothesis, as indicated by the bottom panel of
Figure 5, can also be rejected at the 99\% confidence level.
Thus, at least some amount of triaxiality must be present to
explain the observed lack of nearly circular isophotes.

Once the hypothesis of axisymmetry is abandoned, it requires
two axis ratios, $\beta = b/a$ and $\gamma = c/a$, to describe
the shape of a triaxial galaxy. For a population of triaxial systems,
the distribution of intrinsic axis ratios, $f_t (\beta,\gamma)$, is
no longer uniquely determined by the distribution of observed
axis ratios, $f(q)$. However, I can determine a permissible distribution
of intrinsic axis ratios by assuming a parametric form for $f_t$
and adjusting the parameters to yield the best fit. For instance,
assume that the outer regions of the core galaxies in my sample
have intrinsic axis ratios distributed according to the law
\begin{equation}
f_t (\beta, \gamma) \propto \exp \left[ - {(\beta-\beta_0)^2
+ (\gamma-\gamma_0)^2 \over 2 \sigma_0^2 } \right] \ .
\end{equation}
(There is no particular physical reason to expect $f_t$ to
be an isotropic Gaussian in $(\beta,\gamma)$ space; this
is merely a convenient parameterization.)
The best-fitting set of parameters, as measured by a KS test
to the observed axis ratios of the outer regions of core galaxies,
is $\beta_0 = 0.72$, $\gamma_0 = 0.71$, and $\sigma_0 = 0.037$,
with a KS probability of $P_{\rm KS} = 0.997$. Although the
peak of this distribution is at a nearly prolate shape, it
is also possible to get a good fit with a model in which the
most probable shape is oblate ($\beta_0 = 1$, $\gamma_0 = 0.67$,
and $\sigma_0 = 0.056$, yielding $P_{\rm KS} = 0.87$).

For the inner regions of power law galaxies, where $t_{\rm dyn} =
0.75 {\,\rm Myr}$, the nonparametric estimate for ${\hat f}_o$ is
shown in the middle panel of Figure 6. Although the best fit
to ${\hat f}_o$, given by the solid line, drops below zero for
$\gamma > 0.75$, the oblate hypothesis, given the width
of the error intervals, can't be ruled out at the 91\%
confidence level. The best fit to ${\hat f}_o$ peaks at $\gamma =
0.64$; if the inner regions of power law galaxies are
randomly oriented oblate objects, then very few of them
can be rounder than $\gamma \approx 0.8$. The prolate
hypothesis, as shown in the bottom panel of Figure 6,
is statistically acceptable; it also leads to the conclusion
that few power law galaxies have inner regions rounder than
$\gamma \approx 0.8$.

Since the outer and inner regions of power law galaxies have
a similar distribution of apparent shapes, the nonparametric
estimate of ${\hat f}_o$ for the outer regions of power law ellipticals,
shown as the middle
panel of Figure 7, is similar to that for the inner regions,
shown as the middle panel of Figure 6. For the outer regions,
the best fit peaks at $\gamma = 0.65$, with few galaxies, under
the oblate hypothesis, being rounder than $\gamma \approx 0.8$.
The prolate hypothesis, illustrated in the bottom panel of Figure 7,
is also statistically acceptable for power law galaxies.

The distributions of $q_{\rm out}$ for the core galaxies and
the power law galaxies in my sample are statistically indistinguishable
($P_{\rm KS} = 0.55$). Thus, it is illuminating to adopt the
hypothesis that the shapes of elliptical galaxies in the regime
for $t_{\rm dyn} \approx 50 {\,\rm Myr}$ is independent of whether
their central density cusps are steep or shallow. With this hypothesis,
the samples of $q_{\rm out}$ for core galaxies and power law galaxies
can be combined, to provide better statistics. The combined sample
has 34 galaxies (20 core, 14 power law). For the combined sample,
${\hat f} (q_{\rm out})$ peaks at $q_{\rm out} = 0.75$. The lack of
nearly circular galaxies in the combined sample, along with the
narrower error intervals, means that both the oblate hypothesis
and the prolate hypothesis can be rejected at the 99\% confidence
level. However, the distribution of $q_{\rm out}$ for the combined
sample of core and power law galaxies can be well fit by a
population of triaxial shapes. The best-fitting parametric
function $f_t(\beta,\gamma)$ of the form given in equation (6)
has parameters $\beta_0 = 0.73$, $\gamma_0 = 0.70$, and
$\sigma_0 = 0.095$, yielding 
$P_{\rm KS} = 0.97$. However, acceptable fits are also provided
by models in which the most probable shape is oblate ($\beta_0 = 1$,
$\gamma_0 = 0.64$, and $\sigma_0 = 0.086$, yielding $P_{\rm KS} = 0.78$)
and by models in which the most probable shape is prolate ($\beta_0 =
\gamma_0 = 0.71$ and $\sigma_0 = 0.073$, yielding $P_{\rm KS} = 0.95$).

\section{Summary and Discussion}

In their outer regions, where $t_{\rm dyn} \gtrsim 20 {\,\rm Myr}$,
the core ellipticals in the F97 sample have
a distribution of apparent shapes which is indistinguishable
from that of the power law ellipticals. The scarcity of elliptical
galaxies with circular isophotes in their outer regions implies
that they are probably not all oblate or prolate objects. Due
to the degeneracy of the problem, it cannot be determined, from
the surface photometry alone, whether the majority of elliptical
galaxies are nearly oblate, nearly prolate, or highly triaxial
in their outer regions. (Moreover, even if core ellipticals and power
law ellipticals have an identical distribution of apparent shapes
in projection, this does not demand, although it does permit,
an identical distribution of intrinsic shapes.)

The most intriguing result of my analysis is that in
the region where $0.5 {\,\rm Myr} \lesssim
t_{\rm dyn} \lesssim 3 {\,\rm Myr}$, the core ellipticals are
significantly rounder than the power law ellipticals. In
this regime, the oblate hypothesis is acceptable for
core ellipticals, with a distribution of intrinsic
axis ratios peaking at a fairly round shape, $\gamma \approx 0.86$.
In the very central regions, $t_{\rm dyn} \ll 0.5 {\,\rm Myr}$,
the core ellipticals become flatter again on average, probably
due, in part, to the presence of circumnuclear stellar disks
seen in projection. 

How can we explain the observation that core galaxies are
so nearly spherical in the region where $0.5 {\,\rm Myr} \lesssim
t_{\rm dyn} \lesssim 3 {\,\rm Myr}$? This region is far enough
from the center that circumnuclear disks of stars, gas, and
dust don't strongly affect the isophote shape, but has orbital
times short enough for a central MDO to have completed
chaotic mixing. The observed shapes
of core galaxies are consistent with their being nearly
spherical and nearly oblate in their inner regions, and
more flattened and more triaxial in their outer regions,
just as we would naively expect from the effects of
a central MDO. The shape transition occurs at dynamical
times of $\sim 5 {\,\rm Myr}$, or orbital times of
$\sim 20 {\,\rm Myr}$.
The observation that power law galaxies have the same
distribution of apparent shapes in their inner and outer
regions seems to indicate that the central density cusp
and MDO in power law galaxies is not effective at driving
the stellar orbits in the central region to more spherical
stochastic shapes.

The observed difference in shape between core ellipticals and
power law ellipticals may be linked to a difference in their
origins. Faber et al.\ (1997) have proposed a scenario in which
cores with shallow cusps are the result of the orbital
decay of MDOs accreted during galaxy mergers. Numerical
simulations of the evolution of binary black holes in
the centers of galaxies (\cite{qh97}) has shown that the
galaxy's stellar distribution develops a central core
as stars are ejected from the central cusp by the black hole binary.
The mass of the stellar core which develops is roughly 5 times the
mass of the black hole binary. For the core ellipticals kinematically
modeled by Magorrian et al.\ (1998), the radius $r_c$ at which
$M(r_c) \approx 5 M_\bullet$ ranges from 90 to 7000 parsecs,
with a median value of 600 parsecs. The dynamical time at $r_c$
in these core galaxies ranges from $0.5$ to $26$ million years,
with a median value of $2.3 {\,\rm Myr}$. These dynamical times
are intriguingly close to the dynamical time $\sim 5 {\,\rm Myr}$
at which core ellipticals make the transition from being flattened
and triaxial to being more nearly spherical.

\vskip 0.5in 
This work was supported by NSF grant AST 93-57396. 
Marcella Carollo, Tod Lauer,
and Frank van den Bosch provided me with isophotal data. Rick
Pogge, David Merritt, and an anonymous referee made useful comments.

\clearpage
\begin{figure}
\plotone{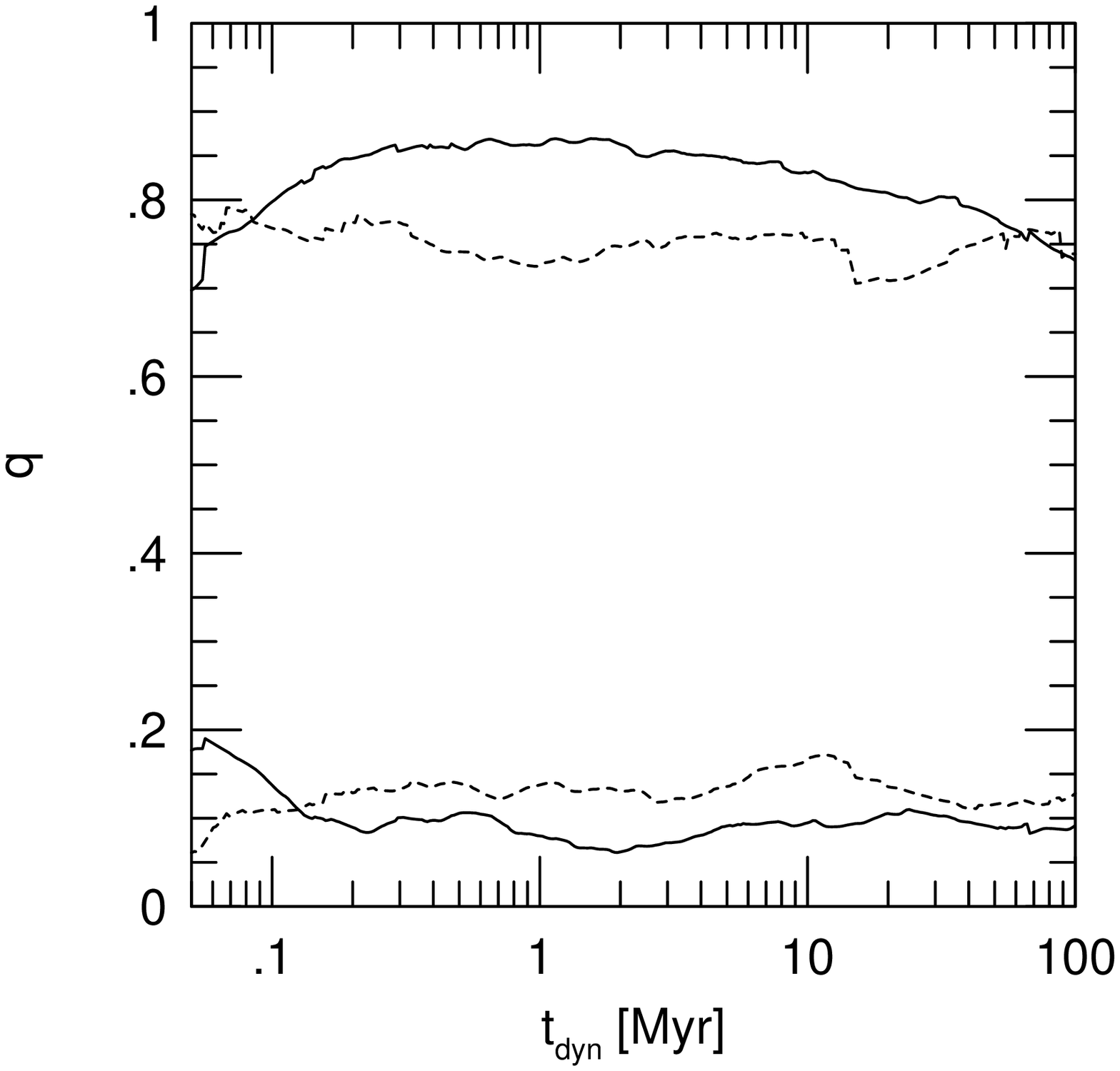}
\caption{The mean (upper lines) and standard deviation (lower lines)
of the axis ratio $q$ as a function of dynamical time, for each
subsample of elliptical galaxies. The solid line indicates the subsample
of core ellipticals, and the dashed line indicates the subsample
of power law ellipticals.
}
\end{figure}

\begin{figure}
\plotone{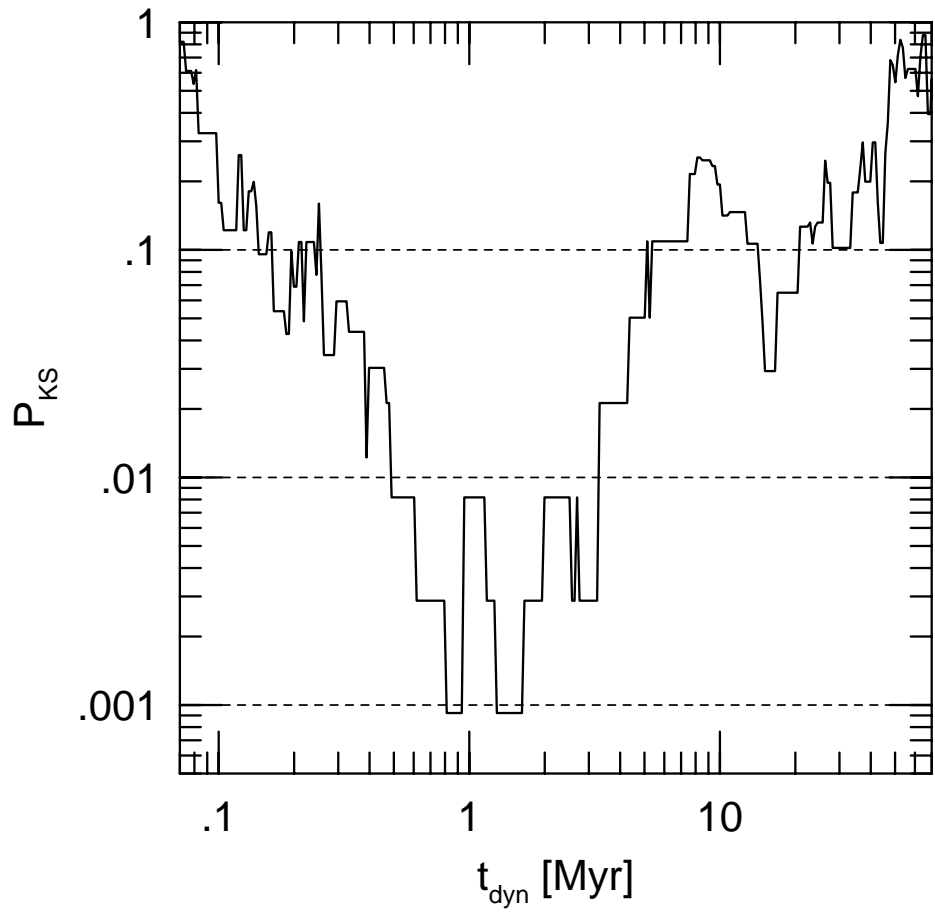}
\caption{The probability yielded by a Kolmogorov-Smirnov test comparing
the distribution of $q$ for the core ellipticals with
the distribution of $q$ for the power law ellipticals.
}
\end{figure}

\begin{figure}
\plotone{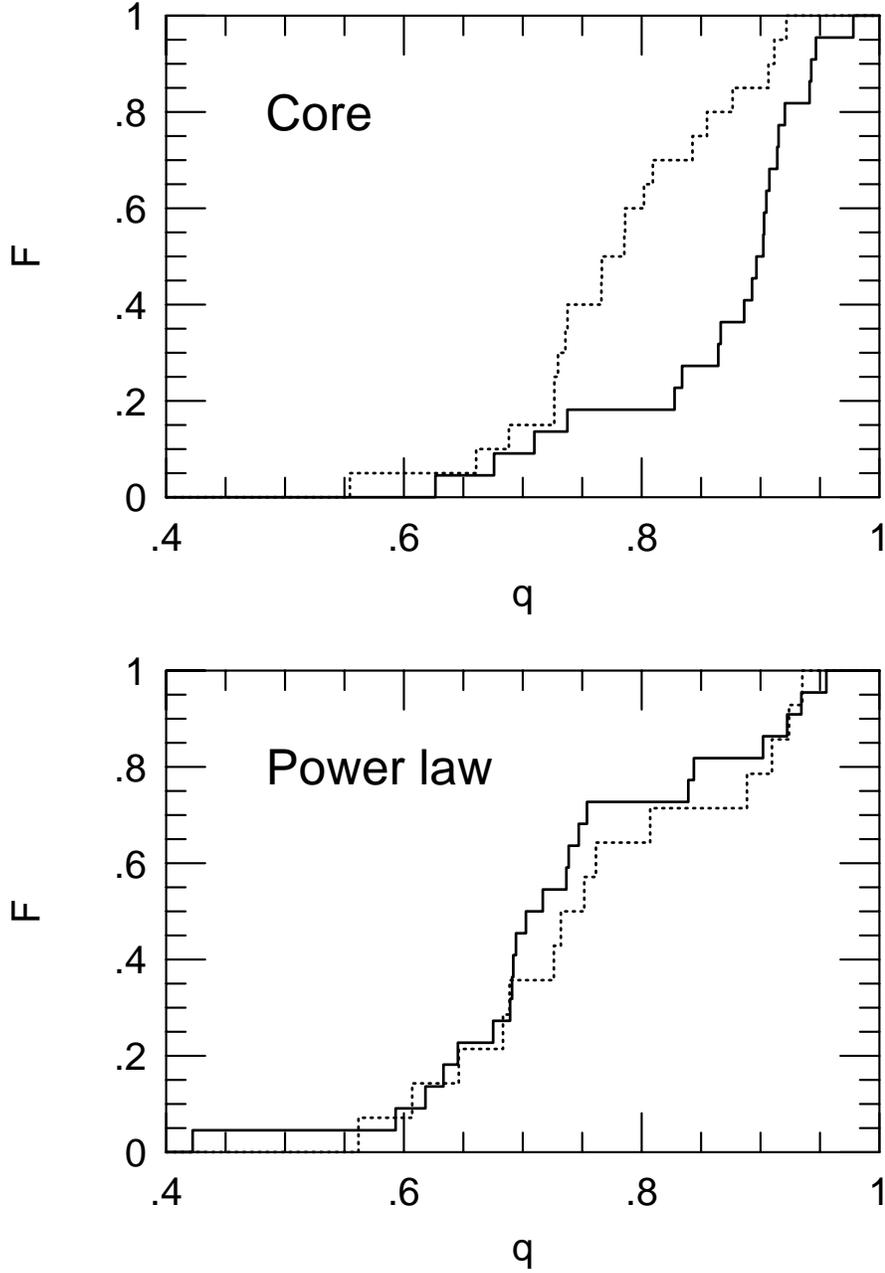}
\caption{{\it Top,} the cumulative distribution function for
the apparent axis ratios of the core galaxies in the sample.
{\it Bottom,} the cumulative distribution function for the
apparent axis ratios of the power law galaxies. In each panel,
the solid line indicates the axis ratios of the $t_{\rm dyn}$ =
0.75 Myr isophote and the dotted line indicates the axis ratios of the
$t_{\rm dyn}$ = 50 Myr isophote.} 
\end{figure}

\begin{figure}
\plotone{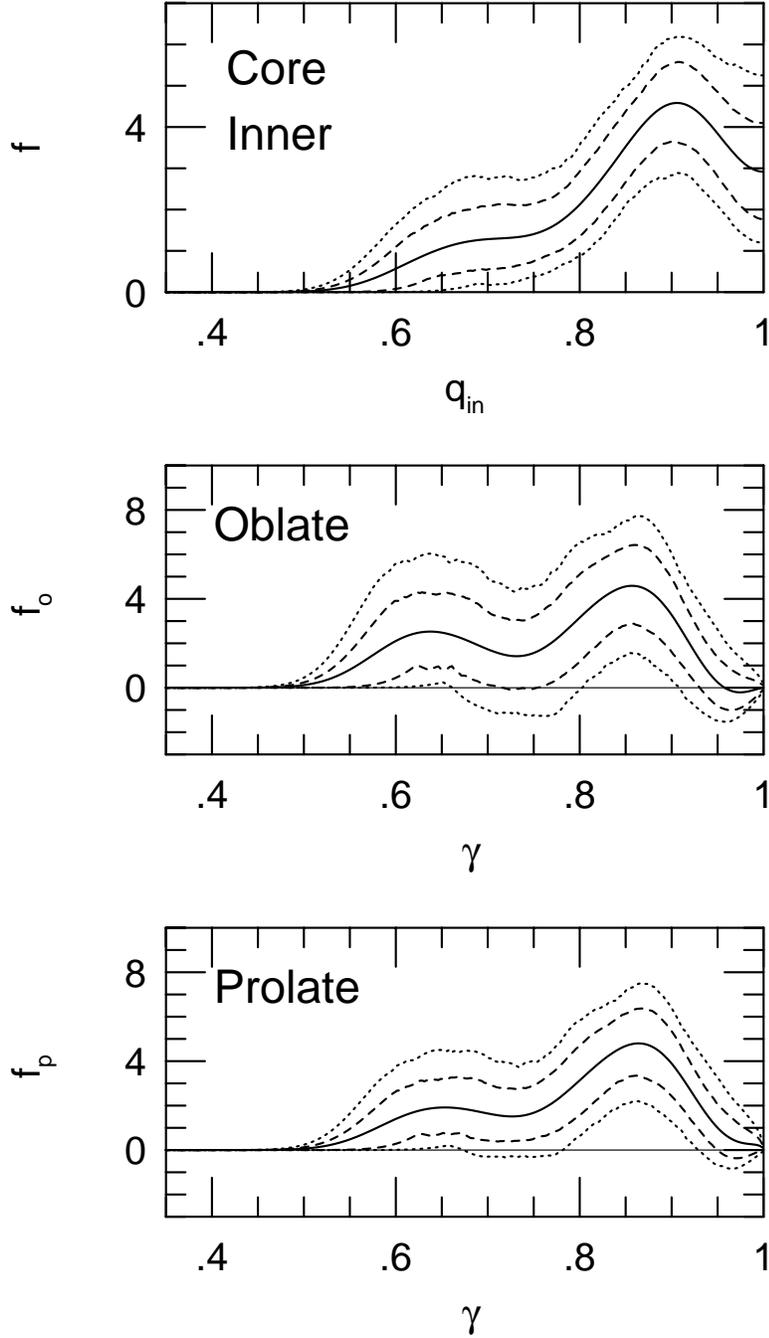}
\caption{{\it Top,} the nonparametric kernel estimate of the
distribution function of axis ratios of the $t_{\rm dyn}$ = 0.75 Myr
isophotes of the core ellipticals. {\it Middle,} distribution
of intrinsic axis ratios, assuming the intrinsic shape is oblate.
{\it Bottom,} distribution of intrinsic axis ratios, assuming
the intrinsic shape is prolate. The solid line in each panel is the
best estimate, the dashed lines are the 80\% confidence band, found
by bootstrap resampling, and the dotted lines are the 98\%
confidence band. The sample contains $N = 22$ galaxies; the
kernel width is $h = 0.044$.}
\end{figure}

\begin{figure}
\plotone{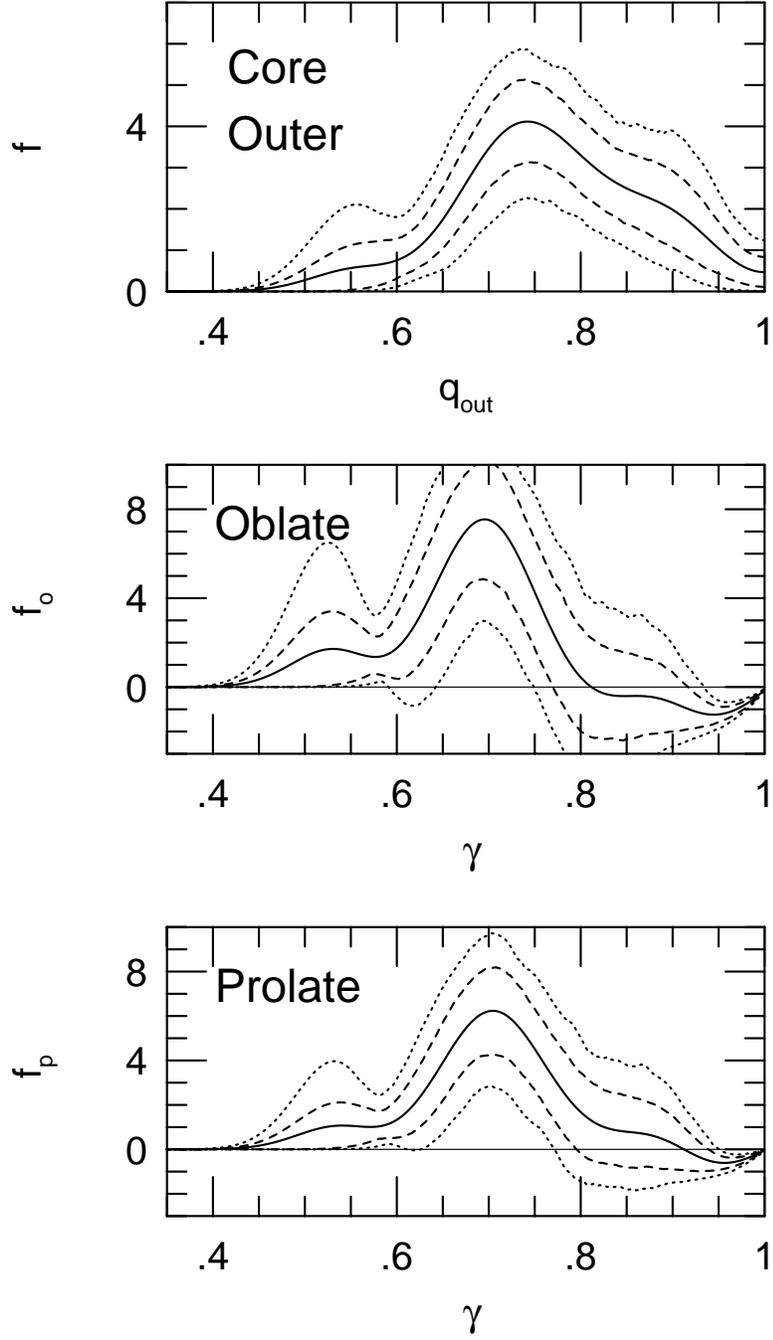}
\caption{As Fig. 2, but for the $t_{\rm dyn}$ = 50 Myr isophotes
of the core galaxies. The sample contains $N = 20$ galaxies;
the kernel width is $h = 0.044$.}
\end{figure}

\begin{figure}
\plotone{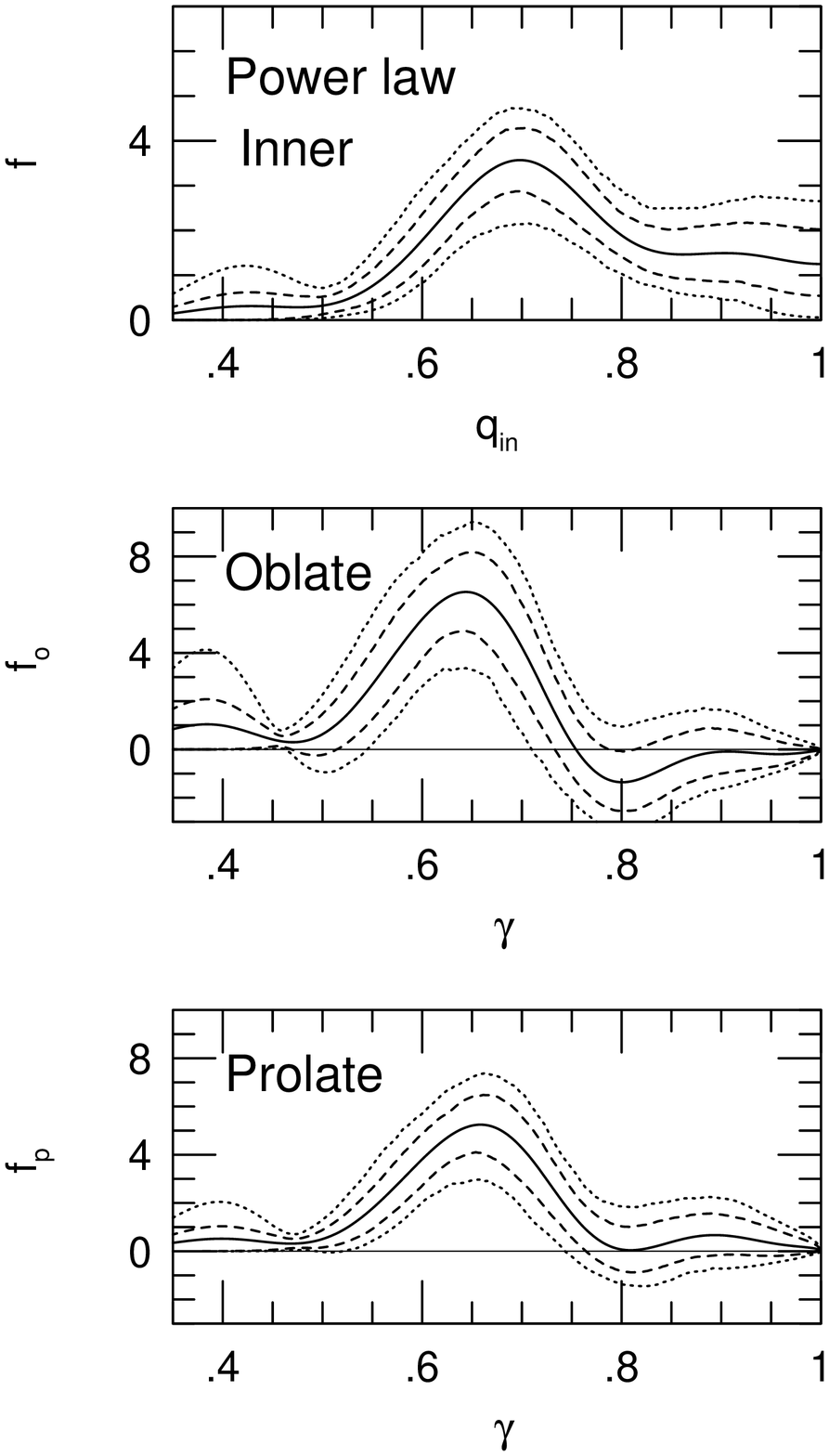}
\caption{As Fig. 2, but for the $t_{\rm dyn}$ = 0.75 Myr isophotes
of the power law galaxies. The sample contains $N = 22$ galaxies;
the kernel width is $h = 0.060$.}
\end{figure}

\begin{figure}
\plotone{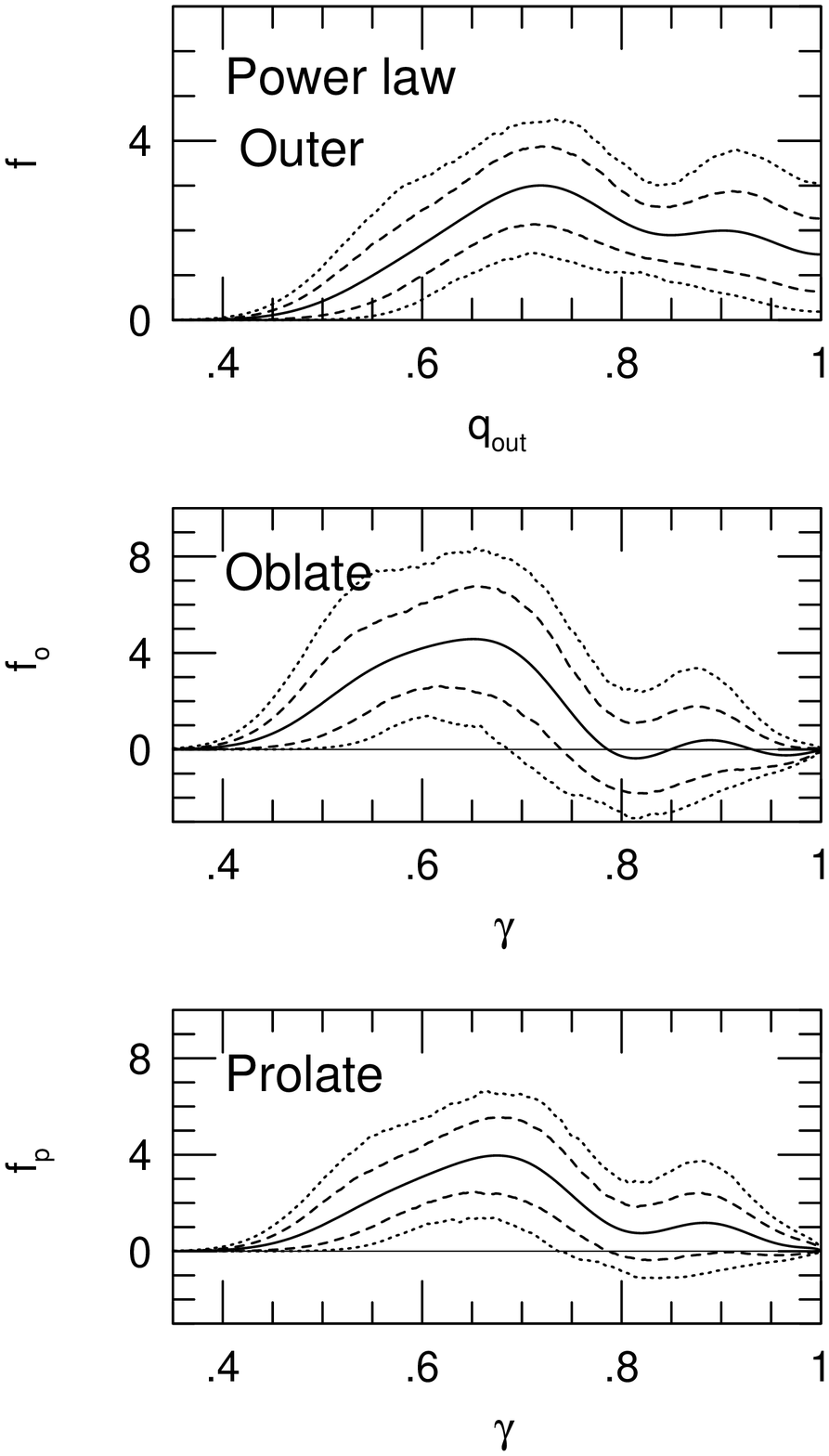}
\caption{As Fig. 2, but for the $t_{\rm dyn}$ = 50 Myr isophotes
of the power law galaxies. The sample contains $N = 14$ galaxies;
the kernel width is $h = 0.061$.}
\end{figure}

\clearpage

\begin{center}
\smallskip
\centerline{\small {TABLE 1}}
\smallskip
\centerline{\small {\sc Apparent Axis Ratios of Core Galaxies}}
\smallskip
\begin{tabular}{lrrcrccc}
\hline
\hline
\multicolumn{1}{c}{Galaxy} &
\multicolumn{1}{c}{Distance} &
\multicolumn{1}{c}{r$_{\rm in}$} &
\multicolumn{1}{c}{q$_{\rm in}$} &
\multicolumn{1}{c}{r$_{\rm out}$} &
\multicolumn{1}{c}{q$_{\rm out}$} &
\multicolumn{1}{c}{Inner} &
\multicolumn{1}{c}{Outer} \\
\multicolumn{1}{c}{} &
\multicolumn{1}{c}{[Mpc]} &
\multicolumn{1}{c}{[arcsec]} &
\multicolumn{1}{c}{} &
\multicolumn{1}{c}{[arcsec]} &
\multicolumn{1}{c}{} &
\multicolumn{1}{c}{isophotes} &
\multicolumn{1}{c}{isophotes} \\
\hline
Abell 1020 & 243.8 & 0.16 & 0.915 &  9.7 & 0.922 & L95 & L99 \\
Abell 1831 & 280.9 & 0.17 & 0.905 &  8.7 & 0.802 & L95 & L99 \\
Abell 2052 & 132.0 & 0.32 & 0.738 & 16.0 & 0.727 & L95 & L99 \\
NGC 720	   &  22.6 & 1.13 & 0.897 & 99.3 & 0.555 & L95 & P90	\\
NGC 1399   &  17.9 & 2.72 & 0.902 & 125. & 0.912 & L95 & C94	\\
NGC 1400   &  21.5 & 1.41 & 0.908 & $\dots$ & $\dots$ & L95 & L85 \\
NGC 1600   &  50.2 & 1.22 & 0.627 & 65.5 & 0.730 & L99 & P90	\\
NGC 2832   &  90.2 & 0.62 & 0.893 & 35.1 & 0.688 & L95 & P90	\\
NGC 3379   &   9.9 & 3.01 & 0.864 & 143. & 0.843 & L99 & P90	\\
NGC 3608   &  20.3 & 1.35 & 0.834 & 65.8 & 0.809 & L95 & G94 \\
NGC 4168   &  36.4 & 0.80 & 0.676 & 51.7 & 0.855 & V94 & C90 \\
NGC 4365   &  22.0 & 1.97 & 0.710 & 110. & 0.766 & V94 & C94	\\
NGC 4472   &  15.3 & 2.70 & 0.914 & 207. & 0.786 & V94 & C94	\\
NGC 4486   &  15.3 & 2.26 & 0.978 & 193. & 0.785 & L92 & C90	\\
NGC 4486B  &  15.3 & 1.74 & 0.828 & $\dots$ & $\dots$ & L95 & $\dots$ \\
NGC 4552   &  15.3 & 2.42 & 0.941 & 110. & 0.877 & L99 & C90 \\
NGC 4636   &  15.3 & 1.04 & 0.943 & 132. & 0.738 & L95 & P90	\\
NGC 4649   &  15.3 & 2.59 & 0.947 & 172. & 0.767 & L99 & P90	\\
NGC 4874   &  93.3 & 0.68 & 0.886 & 27.5 & 0.907 & L95 & P90	\\
NGC 4889   &  93.3 & 0.74 & 0.903 & 35.8 & 0.661 & L95 & P90	\\
NGC 5813   &  28.3 & 1.30 & 0.920 & 73.4 & 0.727 & L95 & P90	\\
NGC 6166   & 112.5 & 0.61 & 0.867 & 25.6 & 0.736 & L95 & L99 \\
\hline
\end{tabular}
\end{center}

\clearpage

\begin{center}
\smallskip
\centerline{\small {TABLE 2}}
\smallskip
\centerline{\small {\sc Apparent Axis Ratios of Power Law Galaxies}}
\smallskip
\begin{tabular}{lrrcrccc}
\hline
\hline
\multicolumn{1}{c}{Galaxy} &
\multicolumn{1}{c}{Distance} &
\multicolumn{1}{c}{r$_{\rm in}$} &
\multicolumn{1}{c}{q$_{\rm in}$} &
\multicolumn{1}{c}{r$_{\rm out}$} &
\multicolumn{1}{c}{q$_{\rm out}$} &
\multicolumn{1}{c}{Inner} &
\multicolumn{1}{c}{Outer} \\
\multicolumn{1}{c}{} &
\multicolumn{1}{c}{[Mpc]} &
\multicolumn{1}{c}{[arcsec]} &
\multicolumn{1}{c}{} &
\multicolumn{1}{c}{[arcsec]} &
\multicolumn{1}{c}{} &
\multicolumn{1}{c}{isophotes} &
\multicolumn{1}{c}{isophotes} \\
\hline
NGC 221  &   0.8 & 14.3 & 0.737 & $\dots$ & $\dots$ & L98 & P93 \\
NGC 596	 &  21.2 & 1.46 & 0.922	& $\dots$ & $\dots$ & L95 & G94	\\
NGC 1172 &  29.8 & 0.90 & 0.839	& 40.7 & 0.726 & L95	& C88	\\
NGC 1426 &  21.5 & 1.20 & 0.694	& 60.5 & 0.646 & L95	& C88	\\
NGC 1700 &  35.5 & 1.33 & 0.739	& 60.0 & 0.683 & L95	& C88 \\
NGC 2636 &  33.5 & 0.46 & 0.955	& $\dots$ & $\dots$ & L95 & $\dots$ \\
NGC 3377 &   9.9 & 3.18 & 0.422	& 97.5 & 0.607 & L95 & P90	\\
NGC 3605 &  20.3 & 0.70 & 0.747	& $\dots$ & $\dots$ & L95 & P90	\\
NGC 4387 &  15.3 & 0.78 & 0.754	& 53.1 & 0.732 & L95	& C90	\\
NGC 4434 &  15.3 & 1.02 & 0.934	& 49.0 & 0.910 & L95	& C90	\\
NGC 4458 &  15.3 & 1.23 & 0.717	& 41.6 & 0.935 & L95	& C90 \\
NGC 4464 &  15.3 & 1.32 & 0.645	& 42.0 & 0.889 & L95	& C90	\\
NGC 4467 &  15.3 & 0.77 & 0.633 & $\dots$ & $\dots$ & L95 & $\dots$ \\
NGC 4478 &  15.3 & 1.31 & 0.689	& 66.4 & 0.924 & V94	& C90 \\
NGC 4551 &  15.3 & 0.91 & 0.675	& 55.3 & 0.762 & L95	& C90	\\
NGC 4564 &  15.3 & 2.07 & 0.703	& 95.2 & 0.562 & V94	& C90 \\
NGC 4621 &  15.3 & 3.01 & 0.618	& 123. & 0.752 & L99	& C90 \\
NGC 4697 &  10.5 & 3.02 & 0.593	& 158. & 0.689 & L95	& P90	\\
NGC 5845 &  28.2 & 1.35 & 0.691	& 41.2 & 0.807 & L95	& R94	\\
VCC 1199 &  15.3 & 0.43 & 0.692	& $\dots$ & $\dots$ & L95 & $\dots$ \\
VCC 1440 &  15.3 & 0.42 & 0.902	& $\dots$ & $\dots$ & L95 & $\dots$ \\
VCC 1627 &  15.3 & 0.42 & 0.844	& $\dots$ & $\dots$ & L95 & $\dots$ \\
\hline
\end{tabular}
\end{center}

\clearpage

\begin{center}
\smallskip
\centerline{\small {TABLE 3}}
\smallskip
\centerline{\small {\sc Kolmogorov-Smirnov Probabilities:}}
\centerline{\small {\sc Comparison of Isophote Axis Ratios}}
\smallskip
\begin{tabular}{lllll}
\hline
\hline
\multicolumn{1}{c}{} &
\multicolumn{1}{c}{} &
\multicolumn{1}{c}{q$_{\rm out}$} &
\multicolumn{1}{c}{q$_{\rm in}$} &
\multicolumn{1}{c}{q$_{\rm out}$} \\
\multicolumn{1}{c}{} &
\multicolumn{1}{c}{} &
\multicolumn{1}{c}{(Core)} &
\multicolumn{1}{c}{(Power law)} &
\multicolumn{1}{c}{(Power law)} \\
\hline
q$_{\rm in}$ & (Core) & 0.0059 & 0.0029 & 0.016 \\
q$_{\rm out}$ & (Core) & $\dots$ & 0.076 & 0.55 \\
q$_{\rm in}$ & (Power law) & $\dots$ & $\dots$ & 0.92 \\
\hline
\end{tabular}
\end{center}

\end{document}